\documentclass{pasj00}
\draft
\begin{document}
\SetRunningHead{Takata et al.}{Dusty ERO behind Two Clusters}
\Received{2002/12/07} 
\Accepted{2003/06/10} 
\title{Dusty ERO Search behind Two Massive Clusters}
\author{Tadafumi \textsc{Takata}\altaffilmark{1},
	Nobunari \textsc{Kashikawa}\altaffilmark{2},
	Kouichiro \textsc{Nakanishi}\altaffilmark{3},\\
        Kentaro \textsc{Aoki}\altaffilmark{1},
        Ryo \textsc{Asai}\altaffilmark{2},
        Noboru \textsc{Ebizuka}\altaffilmark{5},
        Motoko \textsc{Inata}\altaffilmark{4},
        Masanori \textsc{Iye}\altaffilmark{2},\\
        Koji S. \textsc{Kawabata}\altaffilmark{2},
        George \textsc{Kosugi}\altaffilmark{1},
        Youichi \textsc{Ohyama}\altaffilmark{1},
        Kiichi \textsc{Okita}\altaffilmark{4},\\
        Toshiyuki \textsc{Sasaki}\altaffilmark{1},
        Yoshihiko \textsc{Saito}\altaffilmark{2},
        Kazuhiro \textsc{Sekiguchi}\altaffilmark{1},
        Yasuhiro \textsc{Shimizu}\altaffilmark{4},\\
        Hiroko  \textsc{Taguchi}\altaffilmark{2},
        and 
        Michitoshi \textsc{Yoshida}\altaffilmark{4}}
\altaffiltext{1}{Subaru Telescope, National Astronomical Observatory of Japan,\\
650 North A'ohoku Place, Hilo, Hawaii 96720, U.S.A.}
\email{takata@naoj.org}
\altaffiltext{2}{Opt/IR Astronomy Division, National Astronomical Observatory,\\ 
2-21-1 Osawa, Mitaka, Tokyo 181-8588, Japan}
\altaffiltext{3}{Nobeyama Radio Observatory , National Astronomical Observatory,\\ 
462-2 Nobeyama, Minamimaki, Minamisaku, Nagano 384-1305, Japan}
\altaffiltext{4}{Okayama Astrophysical Observatory , National Astronomical Observatory,\\ 
Kamogata, Asakuchi, Okayama 719-0232, Japan}
\altaffiltext{5}{RIKEN(The Institute of Physical and Chemical Research),\\
2-21 Hirosawa, Wako, Saitama 351-0198, Japan}
%
\maketitle
\begin{abstract}
We performed deep $K'$-band imaging observations of 2 massive clusters, 
MS $0451.6-0305$ at $z = 0.55$ and MS $0440.5+0204$ at $z = 0.19$, for
searching counterparts of the faint sub-mm sources behind these clusters, which
would provide one of the deepest extremely red object(ERO) samples.
Comparing our near-infrared images with optical images taken by the Hubble
Space Telescope and by the Subaru Telescope, we identified 13 EROs in these fields.
The sky distributions of EROs are consistent with the previous results, that 
there is a sign of strong clustering among detected EROs.
Also, the surface density with corrected lensing amplification factors
in both clusters are in good agreement with that derived from previous surveys.
We found 7 EROs and 3 additional very red objects in a small area 
($\sim$ 0.6 arcmin$^{2}$) of the MS $0451.6-0305$ field around an extended 
SCUBA source. Many of their optical and near-infrared colors are consistent 
with dusty star-forming galaxies at high redshifts(z $\sim$1.0--4.0), and 
they may be constituting a cluster of dusty starburst galaxies and/or lensed 
star-forming galaxies at high redshift. Their red $J-K'$ colors and faint 
optical magnitudes suggest they are relatively old massive stellar systems 
with ages($>$300 Mega years) suffering from dust obscuration. 
We also found a surface-density enhancement of EROs around the SCUBA source in the 
MS $0440.5+0204$ field. 
\end{abstract}

\section{Introduction}
Following the commissioning of SCUBA on JCMT(Holland et al. 1999),
sub-mm surveys of distant universe have rapidly increased the number of sub-mm selected
galaxies (Hughes et al. 1998; Eales et al. 1999; Barger et al. 1998;
Smail et al. 1997, 2002a; Chapman et al. 2002a; Fox et al. 2002; Scott et al. 2002).
The advantages of selecting high-redshift objects based on a sub-mm 
wavelength are not only their negative {\it{k}}-correction up to $z \sim 10$
(Blain et al. 2002) but also their high sensitivity to very dusty objects, 
which are believed to be undergoing vigorous star
formation at high redshift [their median redshift is revealed to
be $\sim$2.4(Smail et al. 2002a; Chapman et al. 2003)].
They are one of the candidates of forming giant elliptical 
galaxies based on the current observational and theoretical predictions (Lilly et al. 1999). 
Although a very limited number of optical counterparts have been identified spectroscopically,
they show notable diversity in their morphology (extended and/or merging),
AGN possessions, dust temperatures, and so on (Smail et al. 2002a). 
To study the mechanisms of galaxy formations, such as merging and/or
gravitational collapses, we need to clarify the nature and the environments
of these faint sub-mm sources. Clearly, we need a much larger sample of faint sub-mm sources 
than we currently have. It also helps for comparing their features with
nearby analogous objects, such as ultra-luminous IRAS galaxies and/or
some other dusty populations detected by ISO, MAMBO, etc. 
These populations might be missed by the popular ``Lyman Break galaxy surveys"
[there are only few cases of Lyman Break galaxies with sub-mm detection
(e.g. Chapman et al. 2002c)]. On the other hand, as firstly reported by 
Cimatti et al.(1998), there is some evidence that many of the sub-mm sources are 
associated with EROs(extremely red objects)(Cimatti et al. 1998; Blain et al. 2002; Smail et al. 2002a; 
Ivison et al. 2002; Wehner et al. 2002; Frayer et al. 2003).
EROs have red colors, which is consistent with elliptical galaxies at high redshift, and 
they are separated into two categories of dusty starburst population and passively 
evolved stellar systems. The former are believed to be young high-redshift objects with 
a large star-formation rate, and their features are consistent with forming elliptical 
galaxies based on theoretical predictions. 
Therefore, searching for EROs associated with faint SCUBA sources is one of the most effective 
methods for detecting obscured (dusty) star-forming galaxy at high redshift, though it may be
limited only for z$\leq$3, because of large extinction of their optical(rest UV) light by 
their internal dust. There have been several studies (Smail et al. 1999; Ivison et al. 2000; 
Frayer et al. 2000; Barger et al. 2000; Lutz et al. 2001; Frayer et al. 2003) which identified the
ERO counterparts of the sub-mm sources. Furthermore,
Ivison et al.(2000) succeeded to obtain the redshifts of SCUBA sources
at $z \sim 2.56$ and $z \sim 2.2$ in the A 1835 region, and there are
some lower redshift samples with spectroscopically determined redshifts(e.g. Smith et al. 2001).
The lack of a suitable sample is partly because of their optical faintness.
Thus, it is needed to search for ERO counterparts to those sub-mm sources that are optically bright enough 
for detailed spectroscopic studies. 
\par
Using the gravitational lensing effect as a natural telescope, which
amplifies the brightness of distant objects behind the lens, 
is one of the most powerful methods for obtaining
detailed information on intrinsically faint objects.
There have been many successful results in observing various types of high
redshift galaxies using this effect, such as Lyman break galaxies(Pettini et al. 2000;
Franx et al. 1997), faint SCUBA sources(Ivison et al. 2000; Ivison et al. 2001; Ledlow et al. 2002) 
and EROs(G.P. Smith et al. 2002). The advantages of using the 
lensing effect is not only its photometric, but also spatial magnification, which will
provide enlarged images of smaller distant objects, corresponding to a very
high effective spatial resolution, which could not be achieved by the usual ground based
observations. By using these effects, we could investigate the detailed structures and 
also the environmental status of distant objects.
\par
In order to search for bright counterparts of SCUBA sources and to
investigate their surrounding environments in detail, we performed ERO search observations
behind 2 massive clusters of galaxies, MS $0451.6-0305$ ($z = 0.55$) and 
MS $0440.5+0204$ ($z = 0.19$).
These clusters were discovered in the Einstein Medium Sensitivity Survey(EMSS)(Gioia et al. 1990), 
and were covered by a SCUBA mapping observation by Chapman et al.(2002a) and Smail et al.(1998; 2002a).
\par
In the next section, we describe our data collection, including our
observations. In section 3, we address the result of our imaging
surveys and the various features of the candidates of the SCUBA source
counterparts. A discussion is given in section 4. In this paper we use the cosmological
parameters $H_{0} =$60.0 km s$^{-1}$ Mpc$^{-1}$, $\Omega_M = 0.3$, and $\Omega_{\Lambda} = 0.7$.
All magnitudes without any notice are given in Vega magnitude. We used the values in Fukugita et al. (1996) 
for converting between the AB and Vega scale magnitudes. 

\section{Observation, Data, and Data Reduction}
\subsection{Optical Data}
\subsubsection{HST data}
Deep optical imaging data for both clusters taken by WFPC2 
through a F702W filter were retrieved from the HST 
\footnote[1]{Based on observations made with the NASA/ESA Hubble Space Telescope, obtained from the data archive at the Space Telescope Science Institute. STScI is operated by the Association of Universities for Research in Astronomy, Inc. under NASA contract NAS 5-26555.} archive system at STScI. 
The total exposure times for each cluster were 10400 seconds for MS $0451.6-0305$ (Program ID 5987) 
and 22200 seconds for MS $0440.5+0204$ (Program ID 5402), respectively. We used a post-calibrated data and 
combined them into one image by a drizzling technique implemented 
in `dither' package in STSDAS, running in the IRAF
\footnote[2]{IRAF is distributed by the National Optical Astronomy Observatories,
which are operated by the Association of Universities for Research in Astronomy, 
Inc., under cooperative agreement with the National Science Foundation.} environment 
(Fruchter, Hook 2002). The depths for both images are 26.3 and 
26.6 ST-mag($R_{\rm{{ST}}}$)(assuming 2'' diameter aperture with 2$\sigma$ 
detection), respectively. 

\subsubsection{Optical imaging with Subaru Telescope}
Optical imaging using Kron--Cousin's $R_{\rm{c}}$ and $I_{\rm{c}}$ band filters for 
complementing WFPC2 data were performed for MS $0451.6-0305$ 
using the Faint Object Camera and Spectrograph(FOCAS)(Kashikawa et al. 2002) 
at the Cassegrain focus of Subaru Telescope on 2000 December 29, 
during a commissioning observation run of FOCAS. 
Johnson $B$ and SDSS $z'$ imaging observations were performed 
on 2002 November 1 and 2, using Subaru Prime Focus Camera(SuprimeCam)
(Miyazaki et al. 2002) attached to the prime focus of the Subaru telescope. 
All these images, except for $z'$-band image, were taken under good photometric conditions. 
The total exposure times were 3600, 1800, 1800 and 2940 s for the $B$, $R_{\rm{c}}$, $I_{\rm{c}}$ and 
$z'$-bands, respectively. We applied the standard calibration for these data using IRAF
tasks for the FOCAS data, and the software package for SuprimeCam data (Yagi et al. 2002).
Photometric calibrations were performed using ``RUBIN 152'' in Landolt's 
standard stars catalog (Landolt 1992) for the FOCAS data, and SA 95 for the SuprimeCam data
(J.A. Smith et al. 2002). 
The limiting magnitudes of the images were 27.3, 25.6, 24.5, and 24.6 mag for the 
$B$, $R_{\rm{c}}$, $I_{\rm{c}}$ and $z'$-bands, respectively, for 1$\sigma$ detection within 2\arcsec diameter.
Photometric errors for each band were estimated as being equal or less than 0.1 magnitude, 
except for the $z'$-band ($\sim 0.2$). 
For a $z'$ photometric calibration, we used images taken by HST/ACS through F850LP filter, 
and confirmed the accuracy of our photometry to be less than 0.2 magnitude. We used the STSDAS 
synphot package to convert from F850LP magnitude to $z'$ magnitude. 
The seeing sizes were 0.6, 0.6, 0.55, and 0.7 arcsec for the $B$, $R_{\rm{c}}$, $I_{\rm{c}}$ and $z'$ band 
respectively. 
\par
We also retrieved Johnson $V$ image taken by SuprimeCam with 3600 seconds exposure from 
SMOKA(Subaru Mitaka Okayama Kiso Archive system)\footnote[3]{http://smoka.nao.ac.jp.}
(Baba et al. 2002), and reduced it in the standard manner. We performed a photometric 
calibration using standard stars in SA 95 and SA 107(Landolt 1992), and estimated the limiting 
magnitude (2'' aperture with 1$\sigma$) to be 26.7. We checked the catalog 
with published $V$ magnitude data of the same field (Stanford et al. 2002), and estimated 
the errors to be within 0.1 magnitude. 

\subsection{Near-Infrared Imaging}
$K'$-band imaging observations were performed on 2000 November 17 and 18, 
and $J$-band imaging observation was made on 2002 January 6, using the Cooled Infrared Spectrograph 
and Camera for OHS(CISCO)(Motohara et al. 2002) attached to the Nasmyth focus of Subaru Telescope. 
Both clusters' images were taken by using 8 points dithering of 12 exposures with 20 s each 
for the $K'$-band and of 6 exposures with 40 s for the $J$-band, covering 
about a 2\arcmin.0$\times$2\arcmin.0 field of view. The total exposure times 
were 9220 s in the $K'$-band, and 3600 s in the $J$-band for MS $0451.6-0305$ 
and 6880 s in the $K'$-band for MS $0440.5+0204$. The seeing sizes 
were 0\arcsec.6 to 0\arcsec.8 in the $K'$-band, and 0\arcsec.7 to 0\arcsec.9 in the $J$-band. 
The sky conditions were mostly photometric during the observation in 2000 November, 
but not on 2002 January 6. We used FS10 and FS12 from the UKIRT faint standard 
catalog (Casali \& Hawarden 1992) for photometric calibrations. The depth 
of our observations reached to 22.1 mag in $K'$ and 22.3 mag in $J$ with 1$\sigma$ detection 
within 2'' diameter. For $K'$, we estimated the errors in our photometry 
to be less than 0.1 magnitude, but we could not confirm the error for the J-band using our data set, 
because it was not taken under photometric condition. 
Therefore, we compared the $J$-band magnitudes of our sample with those of a somewhat shallower 
sample by Stanford et al (2002) with a limiting magnitude of about 20.5, and found 
no systematic difference between these two datasets(less than 0.1 mag in r.m.s).  
\par
Astrometric calibrations were performed using USNO-A2.0 catalog in each field. 
For MS $0451.6-0305$, we used the $B$-band SuprimeCam image for obtaining a wider field of 
view and about 900 stars with $r$ magnitudes between 16 to 18 for fitting (for avoiding 
the position measurement error by saturation). After the fitting, we applied the 
measured positions of objects in the CISCO $K'$ image and refit them. 
We used the HST F702W image for MS $0440.5+0204$ with 10 stars, because it is the widest 
image for this field. The resultant r.m.s in the fittings were 0\arcsec.20 and 0\arcsec.36 
for MS $0451.6-0305$ and MS $0440.5+0204$, respectively. 

\section{Results}
\subsection{Sample Extraction and Completeness}
We used SExtractor (Bertin, Arnouts 1996) for making the $K'$-band selected catalog, 
which reached to a depth of $\sim$ 22.1 magnitude in 2\arcsec diameter aperture.
An eye inspection was also carried out for all extracted sources, and spurious sources 
were eliminated from the sample. 
Photometry of other bands' images was carried out using the ``double image mode'', 
based on the $K'$-band image. For obtaining precise color information, we applied 
Gaussian smoothing to $B$,$V$,F702W,$R_{\rm{c}}$,$I_{\rm{c}}$,$z'$, and $K'$ frames for matching the image 
qualities to that of the $J$-band, which had the worst seeing size. 
We set the detection criteria at 15 contiguous pixels (0''.105 pix$^{-1}$) 
exceeding 1.5$\sigma$ levels, and detected 235 and 164 objects with $K'\leq22.0$ 
for MS $0451.6-0305$ and MS $0440.5+0204$, respectively.  
The following results and discussions are based on the magnitudes with 2\arcsec  
diameter photometry, which corresponds to a linear scale of 
about 7 kpc and 15 kpc for $z = 0.19$ and 0.55, respectively. 
\par
The completeness of our catalog was estimated by adding artificial objects to 
the $K'$ images for each field at empty positions. We used the IRAF package 
artdata for simulating and adding point-like sources as well as 
objects with de Vaucouleurs and exponential profiles (convolved with the 
seeing PSF) onto the images. SExtractor was run with the same detection 
parameters as for the real data. The limiting magnitudes with 80\% completeness for 
each field were 21.5 for MS $0451.6-0305$, and 21.7 for MS $0440.5+0204$, respectively, 
using point sources, and they decreased to about 65\% for objects with 
exponential profiles at these magnitudes. The completenesses at $K'=22.0$ of point-source 
detection were $\sim$ 50\% and $\sim$ 60\%, respectively for MS $0451.6-0305$ and MS $0440.5+0204$. 
\par
We adopted the photometric system from Holtzman et al. (1995) for 
deriving the $R_{\rm{c}}$-band magnitudes from the HST F702W-band data, 
which are suitable for comparing the previous ERO surveys. We assumed the color of Sbc 
galaxies at a redshift of $z = 1--2$ ($V-R \sim 1.1$), which is 
the same manner of transformation adopted by G.P. Smith et al.(2002). 
The systematic uncertainty of this conversion is $\leq$ 0.1, which 
arises from the likely presence of both more and less evolved 
galaxies than the adopted Sbc spectral type. 
We independently checked the observed FOCAS $R_{\rm{c}}$-band magnitudes with these 
transformed magnitudes for the same objects, and confirmed the consistency 
with each other. In the following discussions, we use the $R_{\rm{c}}$-band magnitudes 
which were transformed from the F702W magnitudes. 
\par
figure 1 shows a $K'$ vs $R_{c}-K'$ diagram for the 2 fields with lines of 
$R_{\rm{c}}-K' = 5.3$ and 6.0, which are our used ERO selection criteria. 
Most of the objects are concentrated around $R_{\rm{c}} - K' \sim 3--4$, which 
is consistent with the color of early-type galaxies of cluster members at the 
corresponding redshifts and is also consistent with the results of shallower observations 
by Stanford et al.(1998) for MS $0451.6-0305$. 
\par
There are 2(8) EROs in the MS $0451.6-0305$ region with $R_{c}-K' \geq$ 6.0(5.3), 
and the number density is 0.5(2.0) arcmin$^{-2}$ down to $K'=$20.8
[it is the limit of $R_{\rm{c}}-K'>5.3$ with $R_{\rm{c}}<26.1$ (R$_{\rm{ST}}<26.5$) sample], 
simply dividing the surveyed area (4 arcmin$^{2}$) without any correction for lens 
amplification. When limiting the magnitude with $K'=$20.0, the numbers 
are reduced to 2(5). 
For MS $0440.5+0204$, there are 0(5) EROs with $R_{\rm{c}}-K' \geq$ 6.0(5.3), 
and the uncorrected number density is 0.0(1.25) arcmin$^{-2}$ down 
to $K'=$21.0.  When limiting the magnitude with $K'=$20.0, the numbers 
are reduced to 0(3). 
\par
figure 2 shows 3-color ($R_{\rm{c}}$,$I_{\rm{c}}$,$K'$) image of MS $0451.6-0305$ with 
2\arcmin $\times$2\arcmin field of view. There are some red objects which are 
bright in the $K'$ band. 
figure 3 shows the sky distributions of EROs in MS $0451.6-0305$ and MS $0440.5+0204$, 
respectively for showing clustering properties in each field. 
It is clear that both in the MS $0451.6-0305$ and the MS $0440.5+0204$ regions, there is  
a non-uniform trend in the sky distribution of EROs and clustering in southern part of  
MS $0451.6-0305$, while near the image center in MS $0440.5+0204$. 

\subsection{Counterparts of SCUBA Sources}
\subsubsection{SCUBA sources in MS $0451.6-0305$}
Two sub-mm sources, SMM J $04542-0301$ and  SMM J $04541-0302$, were detected in a survey of 
massive clusters using SCUBA by Chapman et al. (2002a). 
\par
SMM J $04542-0301$ is located at 25\arcsec from the cluster center, and 
elongated to about 1\arcmin long, which is suggested to be a blend of 
multiple SCUBA sources (Chapman et al. 2002a). Its total flux at 850 $\mu$m is 19.1 mJy, 
and the magnification factor by cluster lensing is estimated to be 4.5. 
There is a constraint of the redshift for this source, as at $z \geq$ 2.3 using 
the `sub-mm' vs `cm' spectral index (Yun, Carilli 2002). 
 We found 7 EROs and 3 optically bright ($R_{\rm{c}} \sim 23--24$) very red objects($R_{\rm{c}}-K' \sim$ 5) around this 
sub-mm source, as shown in figure 4. 
The photometric parameters for these objects are summarized in table 1. 
``A1'' and ``A4'' are very faint, even in a very deep (10400 s exposure) 
F702W image, and ``A2'' and ``A3'' are not visible in it. 
 We plot $R_{\rm{c}}-K'$ vs $J-K'$ diagrams to perform the diagnostics 
for distinguishing dusty star-forming EROs from passively evolved ones, as suggested 
by Pozzetti and Mannucci (2000) for identifying more plausible counterpart of the SCUBA 
source among those EROs in figure 5, in case they are at $1 < z < 2$. 
We also plotted the color tracks of elliptical and dusty starburst galaxies 
based on the GISSEL 99 model (Buruzal, Charlot 1993) for
calculating the SEDs of each track. The ellipticals are represented by a simple stellar 
population with solar metallicity with Salpeter IMF with an age of 15 Gyrs. 
For dusty starbursts, we used a model with a constant star formation rate over 
1Gyr reddened to $E(B-V) =$ 0.8 using the dust-extinction law by Calzetti et al.(2000). 
We used the criteria suggested by Pozzetti and Mannucci (2000) for distinguishing 
two populations of EROs, which is $(J-K) = 0.34(R-K)+0.19$ with $R_{\rm{c}}-K > 5.3$, 
and found that more than half of the EROs near the SCUBA source (``A1'', ``A4'', ``E'' and ``F'') can be 
categorized as dusty starburst populations. 
They may consist of groups of dusty EROs or some blends of lensed objects, 
because ``A1'' to ``A4'' are well aligned along another arc, which is denoted as ``Blue Arc'' in figure 4. 
It should be noted that if many of those EROs do not inhabit at $1.0 < z < 2.0$ this 
diagnostics are not very effective, because there will be overlaps of the occupation area 
between two classes of objects in color-color spaces. For higher redshift objects, 
it is suggested to use $J-K$ vs $H-K$ diagram in Pozzetti and Mannucci (2000). We 
could not perform this diagnostic because of a lack of $H$-band information 
for our sample. More detailed discussions will be given in the next section. 

\begin{table}
  \caption{Photometric properties of EROs and red objects around SMM J $04542-0301$. The upper limits indicate the  
3$\sigma$ detection level in each band. The magnitudes are given in the Vega scale. All of them are observed magnitudes, 
and were not corrected for the lensing effect.}\label{tab:first}
  \begin{center}
    \begin{tabular}{lrrrrrrrrrrr}
     \hline
      Name & \multicolumn{2}{c}{RA  (J2000)  DEC} & \multicolumn{9}{c}{Magnitudes(2\arcsec aperture) and colors}\\
     \hline
           &    &     & $K'$ & $B$  & $V$  & $R_{\rm{c}}$  &  $I_{\rm{c}}$  & $z'$ & $J$ &  $R_{\rm{c}}-K'$ & $J-K'$ \\
     \hline
     \hline
   A1        &04:54:12.81& $-$03:00:44.2& 19.1    & 27.2$^\dagger$& 26.3$^\dagger$& 25.1$^\dagger$& $>$24.5 & 24.4$^\dagger$& $>$22.3 & 6.0   &$>$3.2\\
   A2        &04:54:12.82& $-$03:00:47.6& 20.1    & $>$27.3       & $>$26.7       & $>$26.0       & $>$24.5 & $>$24.6       & $>$22.3 & $>$5.9&$>$2.2\\
   A3        &04:54:12.74& $-$03:00:59.2& 20.5    & 24.6$^*$      & 23.6$^*$      & $>$26.0       & $>$24.5 & 21.2$^*$      & $>$22.3 & $>$5.5&$>$1.8\\
   A4        &04:54:12.68& $-$03:01:01.3& 19.0$^*$& $>$27.3       & 26.3$^*$      & 25.4$^*$      & 24.4$^*$& 23.7$^*$      & 22.1$^*$& 6.4   &3.1   \\
A4(1\arcsec$\phi$)&           &              & 19.8$^*$& $>$28.8       & 27.9$^*$      & 26.1$^*$      & 25.3$^*$& 24.8$^*$      & 23.0$^*$& 6.3   &3.2   \\
   B         &04:54:13.43& $-$03:00:42.9& 19.4$^*$&    27.3$^*$   & 25.5$^*$      & 24.5$^*$      & 23.6$^*$& 23.2$^*$      & 22.0$^*$& 5.1   &2.6   \\
B(1\arcsec$\phi$) &           &              & 20.1    & 28.4          & 26.7          & 24.8          & 24.4    & 24.1          & 23.1    & 4.7   &3.0   \\
   C         &04:54:12.66& $-$03:01:16.3& 18.6$^*$& 26.2$^*$      & 24.6$^*$      & 23.7$^*$      & 22.8$^*$& 22.4$^*$      & 21.6$^*$& 5.1   &3.8   \\
C(1\arcsec$\phi$) &           &              & 19.3    & 27.8          & 26.0          & 24.2          & 23.6    & 23.4          & 22.7    & 4.9   &3.4   \\
   D         &04:54:12.33& $-$03:01:15.6& 19.4    &    27.0       & 25.9          & 24.4          &  24.1   & 23.6          & $>$22.3 & 5.0   &$>$2.9\\
   E         &04:54:10.92& $-$03:01:24.5& 19.5    &    27.1       & 26.0          & 25.6          & $>$24.5 & $>$24.6       & $>$22.3 & 6.1   &$>$2.8\\
   F         &04:54:11.22& $-$03:01:22.9& 19.7    &    26.8       & 26.6          & 25.5          & $>$24.5 & 23.8          & 22.1    & 5.8   &2.4   \\
   G         &04:54:13.66& $-$03:01:21.9& 20.3    & $>$27.3       & $>$26.7       & $>$26.0       & $>$24.5 & $>$24.6       & $>$22.3 & $>$5.7&2.0   \\
      \hline
      \multicolumn{12}{l}{$\dagger$ Possible another source.}\\
      \multicolumn{12}{l}{* Possible contamination from other sources.}\\
    \end{tabular}
  \end{center}
\end{table}
\par
SMM J $04541-0302$ is unresolved in the SCUBA map at 65\arcsec from 
the cluster center, and lens amplification factor is 2.6 (Chapman et al. 2002a). 
The flux at 850 $\mu$m is 16.8 mJy, and the constraint on photometric redshift 
derived from radio index is $z \geq$ 2.2. 
Since this source is out of the field of view of our $J$ and $K'$ images, we can search 
for a counterpart only in optical $B$, $V$, $R_{\rm{c}}$, $I_{\rm{c}}$, and $z'$ images taken by FOCAS and SuprimeCam. 
In these images, a relatively bright galaxy, with $R_{\rm{c}} = 22.2$ with colors of $B-V=1.3$, $B-R_{\rm{c}}=3.0$, 
$R_{\rm{c}}-I_{\rm{c}}=1.0$ and $I_{\rm{c}}-z'=0.55$, is at nearly the SCUBA detection center. 
It has red colors, thus indicating a certain amount of dust obscuration. 
The SED fitting and photometric redshift estimate were carried out with hyperz (Bolzonella et al. 2000) 
with the information from these 5 bands. We had a fitting result with a dusty galaxy at low redshift(around $z \sim 0.6$) 
with $E(B-V) \sim 0.75$. It is a very possible counterpart for the sub-mm source, 
because it is believed to be suffering from large extinction by dust (figure 6). 
This galaxy is already listed in the catalog of the CNOC cluster survey 
(Ellingson et al. 1998), and its $r$ and $g$ magnitudes are 22.53 and 24.23, 
respectively, without information about the redshift. 
Since there is no information on near-infrared nor radio wavebands for the objects, 
it is very difficult to identify a secure counterpart. The possibility of a sub-mm object 
lensed by a bright galaxy invisible in the optical bands, which was suggested by Chapman et al.(2002b), 
can not be eliminated. 

\subsubsection{SMM J $04433+0210$ in MS $0440.5+0204$}
 SMM J $04433+0210$ was discovered in the ``SCUBA Lens Survey'' by Smail et al. (1997, 2002a), 
and its flux at 850 $\mu$m is 7.2 mJy. There is no detected radio flux for this  
source down to the 70 $\mu$Jy level at 1.4 GHz (Smail et al. 2002a). 
An optical counterpart was reported by Frayer et al. (2003) as ``N4'', 
which was detected in their HST/F702W and $K$-band images. 
Their Keck spectroscopy revealed its redshift as being 2.51. 
\par
We measured the $R_{\rm{c}}$ magnitudes for this object as 25.4 and 26.1 for 2\arcsec and 1\arcsec 
apertures, respectively, although there is a certain amount of contamination 
from the neighboring bright galaxy ``N1'', especially for a 2\arcsec aperture (figure 7). 
We also detected this object in the CISCO $K'$ image(``B'' in figure 7); the magnitudes are 
19.5 and 20.3 for 2\arcsec and 1\arcsec apertures, respectively, which are 
in good agreement with the $K$ magnitude in Smail et al. (1999) and Frayer et al. (2003). 
The estimated $R_{\rm{c}}-K'$ color is 5.8 for 1\arcsec diameter magnitudes, and thus it is classified as EROs. 
It should be noted that there are two other EROs (``A'' and ``C'') within 10\arcsec of the sub-mm source. 

\section{Discussions}
\subsection{Lens Model and Surface Density Calculation}
As shown in figure 3 and figure 4, there are concentrations of 
EROs within a small area in both fields. For discussing the surface-density enhancement, 
it is not straightforward to compare our sample with other wide-field ERO surveys, 
because our observation is targeted to the fields of cluster of galaxies.  
The effects by gravitational lensing, such as amplification of the source flux and the accompanying 
distortion of the background sky (source plane), should be taken into account. 
In such a case, lensing models can be used to compute the lens amplification as a function of 
the image plane position. 
\par
We modeled lensing clusters using an elliptical potential with the parameters listed in table 2. 
The lensing inversion was based on the prescription of 
Kormann et al. (1994) and Schneider et al. (1992), allowing 
calculations of the lensing amplification and distortion as a 
function of the position in the image plane. The essential parameters of the model are the redshift 
and the line-of-sight velocity dispersion of the cluster. 
The cluster ellipticity ($f$ in table 2.) is chosen based on 
a measurement of each BCG (Brightest Cluster Galaxy) contour on 
the $K'$-band images. 
\par
The amplification imposed on each background galaxy depends on its redshift(z$_{\rm{source}}$), 
and the redshift of the intervening lens (z$_{\rm{lens}}$). 
We found that cumulative source counts depends weakly on the actual source redshift
by investigating for the case of $z_{\rm{source}}=$1.2, 1.5, 2.0, 2.5, and 3.0, 
at which most of all EROs are expected to lie. Adopting a single value of 
$z_{\rm{source}}$ introduced an uncertainty of at most less than 40\% into the final surface 
density values for the whole field of view in MS $0451.6-0305$, while it was less than 20\% for 
MS $0440.5+0204$. For overcoming these uncertainties, we adopted two single source planes 
of $z_{\rm{source}} =$ 1.5 and 3.0, and used in each cluster model to compute a map of the lens amplification 
as a function of the image plane position for each value of $z_{\rm{source}}$ in each cluster field. 
\par
We first used the amplification maps to de-amplify the image plane flux of each EROs 
in our sample, and hence obtain their source plane $K'$-band magnitudes. Because gravitational 
lensing is achromatic, there is no need to correct the $R_{\rm{c}}-K'$ colors. 
The number of EROs that are brighter than a source plane limiting magnitude can then 
be found by simply counting the number of sources brighter than the limiting magnitude 
in the source plane after correcting for lensing. A simple Poisson uncertainty 
is attached to this value.  
\par
We then calculated the area of the background sky within which EROs were detectable 
in the relevant cluster field in the following manner. An ERO with a source 
plane magnitude of $K'_{\rm{source}}$ will appear in the image plane of the 
relevant cluster with a magnitude brighter than an image plane detection limit 
of $K'_{\rm{det}}$, if it is magnified by a factor greater than 
$\mu_{\rm{min}} = 10^{-0.4(K'_{\rm{det}}-K'_{\rm{source}})}$. 
The area in the source plane within which such a galaxy would be detected 
in that cluster is thus $A$($\mu_{\rm{min}}$), where  $A$($\mu_{\rm{min}}$) is the area 
of the $z_{\rm{source}} = 1.5$ source plane behind each cluster that lies within 
the CISCO field of view, and is magnified by a factor greater than $\mu_{\rm{min}}$(figure 8). 
Therefore, we can derive the surface number density from the dividing number 
of galaxies by $A$($\mu_{min}$). 
\par
\subsection{Corrected Surface Density of EROs}
The derived surface densities for each cluster field are 2.43 arcmin$^{-2}$ 
for $R_{\rm{c}}-K' \geq$ 5.3 in MS $0451.6-0305$, and 2.29 arcmin$^{-2}$ for $R_{c}-K' \geq$ 5.3 
in MS $0440.5+0204$ at $K'=21.6$ for the source plane at $z_{\rm{source}} = 1.5$.   
The cumulative surface densities as a function of the $K'_{\rm{source}}$ magnitude for a sample 
with $R_{\rm{c}}-K' \geq 5.3$ are shown in figure 9. We also plotted the result for 
the 0.6 arcmin$^2$ area in MS $0451.6-0305$ (rectangle surrounded by dotted lines 
in figure 3), in which 7 EROs were detected. 
About 20 and 10\% of the whole survey areas reach to a depth of $K'_{\rm{lim}} \sim 23.0$, 
when we adopt the detection limit as $K'=22.0$. Because of the limit by the depth of the $R$-band 
magnitudes, the depth of our ``ERO samples'' are decreased to $K_{\rm{source}}\sim22.0$. 
Our sample thus provides one of the deepest ERO sample ever reported. 
It is noted that we obtained mostly the same results using other model parameters 
used in Wu et al. (1998) and Williams et al. (1999). 
\par
For a comparison, we overplotted the result of G.P. Smith et al.(2002) and 
Wehner et al.(2002) without the sample of A 2390 in figure 9. Both ERO samples were made 
by targeting lensing cluster fields, and selected by the same color criterion as ours 
($R_{\rm{c}}-K'>5.3$). 
We also overplotted the results by blank-field ERO searches performed 
by Daddi et al.(2000), Roche et al.(2002), and Moustakas et al.(1997). 
G.P. Smith et al. (2002) surveyed lensed EROs behind 10 massive clusters 
of galaxies around $z\sim$0.2 to a depth of $K$ $\sim$ 20.6, and measured the 
surface density of 1.2 arcmin$^{-2}$ for $R_{\rm{c}}-K \geq$ 5.3. They corrected the amplification 
effect from gravitational lensing using their own lens model, and estimated 
the corrected surface densities at $K$ $\leq$ 21.6 as 2.5 arcmin$^{-2}$. 
The values are very similar to our result. 
As shown in figure 9, there is little difference between both samples 
at the overlapping magnitude ranges, and can be considered to be within error bars 
for $R_{\rm{c}}-K' \geq 5.3$ samples. 
\par
For a surface-density enhancement evaluation, we also calculated corrected surface densities in the 
areas which are shown in figure 3 by rectangles with dotted lines. 
In MS $0451.6-0305$ field, 6 EROs in $\sim$0.6 arcmin$^2$ area, and the corrected surface 
density is 46.2 arcmin$^{-2}$ down to $K'_{\rm{source}}=21.0$ at $z_{\rm{source}}$=3.0, or 30.8 at 
$z_{\rm{source}}$=1.5. In MS $0440.5+0204$ field, 3 EROs brighter than $K'=20.6$ are 
within 0.1 arcmin$^{2}$ area, and corrected surface density is 62.5 arcmin$^{-2}$ down to 
$K'_{\rm{source}}=21.0$ at $z_{\rm{source}}$=1.5. 
They are more than 7-times than the average values reported by previous surveys.  
It is the same trend with the sample of G.P. Smith et al. (2002), which indicates that the 
numbers and sky distributions of EROs are very different in each field of view. 
Those are also consistent with the results by other wide field ERO surveys, which suggest the 
clustering properties of EROs (Daddi et al. 2001; Thompson et al. 1999). 
The most different property of our sample from G.P. Smith et al.(2002)'s is that 
most EROs identified in MS $0451.6-0305$ show the colors which are  
common in dusty star-forming galaxies at high redshift(Figure 7), 
which indicate their strong star-formation activities, while most EROs 
are passively evolved objects in Smith et al.'s. 
Daddi et al.(2002) suggested that the strength of clustering is stronger in the old ERO 
population than that of dusty starburst EROs, and our results on the MS $0451.6-0305$ may be a 
very rare case. It may indicate that the deeper is the depth of observation, for 
example down to $K'=20.5$, the more will the ratio of dusty starbursts in ERO population increase. 
This is suggested by Smail et al. (2002b) with a radio-selected ERO sample, and the ratio 
of the dusty population to the passive one is comparable at brighter magnitudes (Cimatti et al. 2002). 
However, we also have to pay attention to the effect of the selection bias in our sample, which is made by 
targeting to lensed SCUBA sources, because it may increase the possibility of 
detecting dusty EROs. 
\par
In the deepest magnitude ranges of our sample, there are deficiencies of EROs 
compared to Wehner et al. (2002), while the uncertainties of ours and of their 
results are very large. This is because both surveys have a very limited region
coverages and also suffer from field variances of ERO numbers. 
\par
\subsection{Lensed Dusty Star-Forming Galaxies around SMM J $04542-0301$?}
\subsubsection{Possible giant extremely red arc}
As mentioned in the previous section, 7 EROs are concentrated within a 0.6 arcmin$^2$ 
sky area, and its corrected surface density is more than 7-times the average value (figure 9). 
For confirming the over density, however we have to take care of the fact that there are EROs 
which are possible to be gravitationally lensed and separated (splitted) into several images
(e.g. ``A1'' to ``A4'' in figure 4). 
These sources are well aligned in parallel to the ``Blue Arc'', which is believed to be 
gravitationally elongated and split. If they are constructing an extremely red arc, the surface 
density will be decreased to 7.7 arcmin$^{-2}$ for $z_{\rm{source}}=1.5$ or 23.1 arcmin$^{-2}$ for 
$z_{\rm{source}}=3.0$ at $K'_{\rm{source}}<21.0$; however it is still a density enhancement. 
\par
For discussing the possibility of this lensing effect, we used the elliptical lens model, 
which was applied to a surface number count. The lens model also gives us a possibility to estimate  
the redshifts of lensed objects without any spectroscopic data, and may be able to set  
constraints on the spatial distribution of some EROs. We compared the critical curves on the image plane 
for different source plane redshifts, as shown in figure 10. We also overplotted 
the contour map of soft X-ray(0.2--1.5 keV) image of Chandra observation
\footnote[4]{Based in part on data retrieved from the Chandra Data Archive(CDA), 
a part of the Chandra X-ray Observatory Science Center(CXC) which is operated for 
NASA by the Smithsonian Astrophysical Observatory.}
with 15000 s exposure\footnote[5]{Because of spurious counts on a soft band 
image, we did not use the deeper (45000 s exposure) ACIS-S image for our analysis. 
For hard band image analysis, we used it.} to show the adequacy of our 
lensing model for approximating the cluster mass distribution. 
The images were retrieved from the Chandra archive, and we used CIAO software 
for extracting the soft band photon, combining each exposure and 
smoothing the resultant image. We applied 5-pixel Gaussian adaptive smoothing for the image using 
the csmooth task. 
The shapes of the inner X-ray contours and the used mass distribution are in  
good agreement for a first-order approximation of the mass distribution. 
\par
The sources ``A1'' to ``A4'' are very near to the critical curve, especially for 
$z_{\rm{source}}=1.2$ in figure 10. Therefore, it is possible that 
they are forming a large extremely red arc, and are split images of an object at $z\sim1.2$.  
If they are, their colors must be very similar to each other, assuming 
no extinction differences along each light path. 
The colors of ``A1'' and ``A4'' are different, although the optical band magnitudes 
may be contaminated by other sources which are near to these objects. 
There remains a possibility that the colors of ``A2'' and/or ``A3'' will coincide with those of 
``A1'' and/or ``A4'' because ``A2'' and ``A3'' are detected only in the $K'$ band, and a color 
comparison was unavailable. 
Although the estimated redshifts is $z = 1.2$ by our lensing model, 
this is inconsistent with a photometric redshift of $z > 2.3$ derived from the 
radio vs sub-mm index if they are counterparts of the sub-mm source. However it is 
possible that the photometric redshift estimate is not appropriate if the assumed 
dust temperature for the SED model is not correct (Blain et al. 2003). In fact, it is suggested 
that there is a diversity of the dust temperature of sub-mm sources based on a sample 
of sub-mm sources with redshifts (Chapman et al. 2003). 
\par
It should also be noted that X-ray emission can not be completely fitted by a simple elliptical mass 
distribution. It is elongated to north-west direction, which has already been suggested by Donahue et al. (1995) 
using the ROSAT image, and the X-ray contours are uneven(figure 10). 
For a more detailed discussion concerning the reality of a giant extremely red arc, it is necessary to use a lens model 
which is more sensitive to the local mass distribution, like that by Kneib et al.(1996). 
It will be considered in a future coming paper with some redshifts of bright objects and blue arcs. 
\par

\subsubsection{SED fitting for a redshift estimation}
For investigating the spatial clustering properties of these EROs and red galaxies 
near SMM J $04542-0301$, we used a SED fitting for deriving the photometric redshifts for 6 objects 
among those (extremely) red objects (``A4'', ``B'', ``C'', ``D'', ``E'', and ``F''), which had confirmed 
magnitudes in several bands. Because the optical bands' lights of ``A1'' may be coming from a 
physically different source (optical centroids are more than 1\arcsec apart from $K'$ image's centroid), 
we eliminated ``A1'' from a list of our SED fitting. 
To avoid the effect of contamination from other surrounding sources, 
we used 1\arcsec diameter magnitudes for sources ``A4'', ``B'', and ``C'', while they were 2\arcsec for others. 
We used hyperz for the calculation, by stepping the extinction from $A_{V}=$ 0.0 to 5.0 
using the extinction formula by Calzetti et al.(2000), and seeked the 
best solution in the range of $0.0 \leq z \leq 10.0$.  
As shown in figure 11, many objects, except for ``D'' and ``F'', were fitted by 
SEDs of galaxies at high redshift($z \geq 2.6$), which are consistent with 
the photometric redshift from sub-mm to radio index($z_{\rm{photo}}\geq2.3$), although with relatively small 
extinction[$E(B-V) \sim 0.15--0.30$]. These extinction values are equal to, or slightly higher than,  
the median of those of Lyman break galaxies at $z\sim3$ (Steidel et al. 1999; Shapley et al. 2001). 
Their $R-K'$ colors are more than 5.0, and thus redder than the sub-mm source Lyman break galaxy ``W-MMD11''
(Chapman et al. 2002c). They also have red $J-K'$ colors ($>2.3$), and their properties are 
similar to those of a red massive galaxy population at high redshift($2<z<4$), revealed by 
the Faint InfraRed Extragalactic Survey(FIRES; Franx et al. 2003; van Dokkum et al. 2003). 
This red color can only be explained by a composite of the stellar ages 
with $\geq$300 Myr and a certain amount of obscuration by dust (Franx et al. 2003). The ages 
for our red sources best fitted by our analysis range from 0.5 to 1.2 Gyr. Daddi et al. (2003) 
suggested that their strong clustering at $2<z<4$ and our result can show the property. 
The derived extinction for ``D'' is large (0.89), and it is a plausible nearby candidate of 
a sub-mm source. 
Since the magnitudes of ``A4'' may be contaminated by neighboring foreground objects and/or contribution 
from diffuse emission by the cluster, the fluxes in the optical bands may be mainly coming from other sources. 
Therefore, it can have redder colors than those listed in table 1, and we could not eliminate the 
possibility of ``A4'' as being a counterpart of a sub-mm source. 
If so, its redshift can be higher, and may be the same as those of ``B'' and ``C''. 
If ``A1'' to ``A4'' are constructing an extremely red arc, it is another plausible candidate of a sub-mm source, 
and they may have a spatial concentration with ``B'' and ``C''. 
\par
It is noted that the results of photometric redshifts using different extinction models, 
like Fitzpatrick(1986) for Large the Magellanic Cloud or Prevot et al. (1984) and 
Bouchet et al. (1985) for the Small Megellanic Cloud, provided slightly higher redshifts. 
\par
By combining the result of our analysis, we can suggest one possibility that 
this extended sub-mm flux is coming from a high-redshift object, which must be lensed 
by a foreground cluster, and there may be a contribution from a lower redshifted ($z \sim 0.5$) 
dusty object, ``D''. 
Since there is some evidence that Lyman break galaxies are not main contributor 
to the cosmic sub-mm background radiation (Adelberger, Steidel 1999; Webb et al. 2002), 
``B'', ``C'', and ``E'' may be eliminated from the counterpart candidates; however, they 
may be spatially clustering near to the sub-mm emission. 
For secure counterpart identification and confirming the speculation of their clustering properties, 
it is essential to perform a deep radio interferometric observation and spectroscopic follow-up 
observations in optical or near-infrared wavelength. 
\par
Using Chandra hard (4--8keV) X-ray images, which was produced from 
the same data as that for figure 10, we checked the coincidences of X-ray 
detections around sub-mm sources in the MS $0451.6-0305$ region, although no counterparts were 
identified. This is consistent with the result of previous studies because about 
50000-s exposure by Chandra is not sufficient for detecting the X-ray flux from 
a large amount of sub-mm sources (Alexander et al. 2003; Waskett et al. 2003).  
\par

\begin{table}
  \caption{Parameters of clusters used in the lensing model. }\label{tab:second}
  \begin{center}
    \begin{tabular}{crrrrrr}
     \hline
     \hline
      Name &  RA(J2000.0) & DEC(J2000.0)  &  $z$  &  $\sigma_{\rm{los}}$(km s$^{-1}$)  &  Reference & $f$ \\
     \hline
      MS $0451.6-0305$ & 04:54:10.8 & $-$03:00:51.6 & 0.55 & 1354 & Ellingson et al.(1998) & 0.9 \\
      MS $0440.5+0204$ & 04:43:09.9 & +02:10:19.8 & 0.19 & 872  & Gioia et al. (1998) & 0.95 \\
      \hline
    \end{tabular}
  \end{center}
\end{table}

\section{Summary}
We performed deep near-infrared and optical imaging of two massive clusters of galaxy fields. 
By combining this data with photometric information taken from the deep WFPC2/F702W image, we identified 
EROs in each field, 8 in MS $0451.6-0305$ and 5 in MS $0440.5+0204$ with $R_{\rm{c}}-K'\geq$ 5.3,  
and $K'<$22.0, in a 2\arcmin$\times$2\arcmin field of view.
Our sample provide one of the deepest sample of EROs ever reported. 
The surface density of our whole-area sample shows good agreement with other previous 
samples at the bright part. At the faint end, the number of EROs is slightly deficient compared 
to the previous work, though it is within the error bars. 
\par
We identified 7 EROs ($R_{\rm{c}}-K'\geq$5.3) and 3 red objects ($I_{\rm{c}}-K'>4.0$) with $K'\leq$ 20.6 
within 0.6 arcmin$^{2}$ of an extended SCUBA source in the MS $0451.6-0305$ field. 
Our investigation using a lens model provided a value of surface density of EROs of  
$\sim$46 arcmin$^{-2}$ for $K'_{\rm{source}} < 21.0$, showing a surface-density enhancement of 
extremely red galaxies, even after a correction of the lensing effect. 
It is possible that 4 sources, named ``A1'' to ``A4'', constitute 
an extremely red-giant arc, and that the best-fit redshift by our model is $z\sim1.2$. 
The photometric redshift for a sub-mm source, derived based on the sub-mm to radio index, 
does not give consistent redshift. 
We also investigated the optical and near-infrared photometric features for the 6  
red sources, and identified that 4 of them have SEDs which are typical of high-redshift 
objects. Their extinctions are typical, or slightly higher, values compared to 
those of Lyman break galaxies at $z\sim3$. Our redshift estimation for ``A4'' based 
on an SED fitting is consistent with that from sub-mm to radio index, and it is possible 
to consider that ``A1'' to ``A4'' are lensed extremely red sources at $z>2.5$, and a 
counterpart of an extended sub-mm source. 
One source with severe extinction is identified at a lower redshift($z\sim0.5$), 
and is one of the candidates of contributors to the extended sub-mm flux. 
For revealing their clustering properties and the relation with extended sub-mm emission, 
a spectroscopic follow-up observation in the optical and/or near-infrared wavelength is essential. 
\par
\vspace{4mm}
This study is based mainly on data collected at Subaru Telescope, and that obtained from the data archive 
at the Astronomical Data Analysis Center, which is operated by the National Astronomical Observatory of Japan. 
The authors thank an anonymous referee for valuable comments and suggestions for improving the manuscript. 
We are grateful to all Subaru staff members for their enormous 
help during our observations. T.T. thanks to Dr. Kentaro Motohara and 
Dr. Chris Simpson for their enormous help during the observation and data 
reduction, Dr. Yoshitomo Maeda for giving instructions on the CIAO software 
for reducing the Chandra image. T.T. also thanks the staff of the archive system 
of the Space Telescope Science Institute for their help in retrieving the HST data. 
This work is partly supported by the Ministry of Education, Culture, Sports, 
Science and Technology in Japan under Grant No. 12740126. 
K.N. has been financially supported by a JSPS Fellowship. 
\par

\newpage
\begin{center}
Figure Captions\\
\end{center}
\begin{itemize}
  \item{Figure 1. (a) $K'$ vs $R_{\rm{c}}-K'$ plot for MS $0451.6-0305$. Color of $R_{\rm{c}}-K'=$ 5.3 and 
6.0, which represent those for EROs, are shown by the dashed and dash-dotted line, 
respectively. The line indicating the 2$\sigma$ limit of the $R_{\rm{c}}$-band image is also 
shown. (b) Same figure for MS $0440.5+0204$.   }
\\
  \item{Figure 2. True-color $R_{\rm{c}}I_{\rm{c}}K'$ image of MS $0451.6-0305$ (using $R_{\rm{c}}$ and $I_{\rm{c}}$ data by FOCAS and 
a $K'$ image by CISCO). This is a 2\arcmin$\times$2\arcmin field of view. We identify several very red 
objects mainly detected in $K'$ image. }
\\
  \item{Figure 3. Sky distributions of EROs in the regions of (a) MS $0451.6-0305$ and 
(b) MS $0440.5+0204$. Filled red circles, filled orange stars, filled orange circles 
represent ERO with $R_{\rm{c}}-K' \geq$ 6.0, ones with 5.3 $\leq R-K' \leq$ 6.0 with and without 
upper limit value of $R_{\rm{c}}$ magnitudes, respectively. Triangles represent other sources 
detected in $K'$-band image for those fields. Regions encircled by dotted lines are showing 
the area we used for surface density calculations. The dashed line in (b) is showing the 
edge of F702W image. Central coordinates in J2000.0 for both fields are denoted at the 
top of the panels.}
\\
  \item{Figure 4. $K'$ image of the area around SMM J $04542-0301$ with an overplotted 
SCUBA contour map by Chapman et al. (2002a). The contour levels represent the 
2$\sigma$ to 4$\sigma$ detection levels. EROs around the SCUBA source are 
represented as ``A1'' to ``A4'' and ``E'', ``F'', and ``G'' with red rectangles, and 
optically bright red objects($R_{\rm{c}}-K' \sim$ 5) ``B'', ``C'', and ``D'' are also shown 
along with purple ones.  
The arc, which is clearly visible in the optical images, is denoted as ``Blue Arc''.}
\\
  \item{Figure 5. $R_{\rm{c}}-K'$ vs $J-K'$ diagram for the objects with red colors in the MS $0451.6-0305$ region. 
The dotted lines indicate the boundary between passively evolved galaxies (left side) and dusty 
starbursts (right side) with an extremely red color($R_{c}-K' = 5.3$). The objects which are identified 
near the extended SCUBA source are denoted as in figure 4. The curves indicate the track of 
elliptical galaxies(solid line) and dusty starbursts(dashed line) with a redshift range of 1 to 4
(redshifts 1.0, 2.0, 3.0 are denoted for each track). The details can be referred to in the text. }
\\
  \item{Figure 6. FOCAS $R_{\rm{c}}$ and $I_{\rm{c}}$ band and SuprimeCam $B$ and $z'$ images around SMM J $04541-0302$ in MS $0451.6-0305$.
The field sizes are 30$\times$30 arcsec$^2$. A contour map of SCUBA detection is overplotted on the $R_{\rm{c}}$ band image with the 
same levels as in figure 4.}
\\
  \item{Figure 7. F702W and $K'$ images around SMM J $04433+0210$ in MS $0440.5+0204$. 
EROs are denoted as ``A'', ``B'', and ``C''. ``B'' is the optical counterpart of 
the sub-mm source(``N4'' in Frayer et al. 2003) at $z=2.51$. Field sizes are 30$\times$30 arcsec$^2$. 
A bright galaxy neighboring ``B'' is ``N1'' in Smail et al. (1999). 
The overplotted contours in $K'$ band image are 850 $\mu$m map in Smail et al. (1998). }
\\
  \item{Figure 8. Cumulative area of the source plane in our survey at $z = 1.5$  
that experiences magnification greater than $\mu$. }
\\
  \item{Figure 9. Amplification-corrected surface densities of EROs for surveyed regions. 
The corrections were applied assuming a single source plane at $z_{\rm{source}}=1.5$, as 
discussed in the text. The triangles indicate the cumulative surface densities for 
MS $0440.5+0204$, the filled pentagons and circles for MS $0451.6-0305$ within a small region as 
shown in figure~\ref{fig:fig3}, and for the whole region, respectively. The open circles show the results of 
G.P. Smith et al.(2002), and the open squares represent those by Wehner et al.(2002). 
We also plotted the results of field ERO searches by Daddi et al. (2000), Moustakas et al. (1997), 
and Roche et al. (2002). We plotted them without any correction for the data with 
$R_{\rm{c}}-K > 5.0$ or $I-K > 4.0$. The uncertainties for our data are 1$\sigma$, 
based on the Poisson distribution. }
\\
  \item{Figure 10. $K'$ image of MS $0451.6-0305$ with the contour of soft (0.2--1.5 keV) 
X-ray image taken by Chandra with a 10 ks exposure overplotted. The levels show 
between 1.0$\times$10$^{-8}$ to 5.0$\times$10$^{-8}$ photons cm$^{-2}$s$^{-1}$pix$^{-1}$ with the 
same interval. The critical curves for the source at $z_{\rm{source}}=$1.2, 1.5, 2.0, 2.5, and 
3.0 are also shown as dotted blue, light blue, green, orange and red lines, respectively. 
The objects denoted in figure 4 are all shown encircled by squares (Red rectangles represent EROs.).}
\\
  \item{Figure 11. Fitted SEDs for 6 (extremely) red objects around SMM J $04542-0301$. }
\end{itemize}

\begin{thebibliography}{}
\bibitem[Adelberder \& Steidel (2000)]{key-1}
   Adelberger, K.L., \& Steidel, C.C. \ 2000, ApJ, 544, 218
\bibitem[Alexander et al. (2003)]{key-2}
   Alexander, D.M., et al. \ 2003 AJ, 125, 383
\bibitem[Baba et al. (2002)]{key-4}
   Baba, H. et al. \ 2002, Rep. Natl. Astron. Obs. Japan, 6, 23 (in Japanese)
\bibitem[Barger et al. (1998)]{key-5}
   Barger, A.J., Cowie, L.L., Sanders, D.B., Fulton, E., Taniguchi, Y., Sato, Y., Kawara, K., \& Okuda, H. \ 1998, Nature, 394, 248
\bibitem[Barger et al. (2000)]{key-6}
   Barger, A.J., Cowie, L.L., \& Richards, E.A. \ 2000, AJ, 119, 2092
\bibitem[Bertin, Arnouts (1996)]{key-8}
   Bertin, E., \& Arnouts, S. \ 1998, A\&AS, 117, 393
\bibitem[Blain et al. (2002)]{key-10}
   Blain, A. W., Smail, I., Ivison, R.J., Kneib, J.-P., \& Frayer, D.T. \ 2002, Physics Report, 369, 111
\bibitem[Blain et al. (2003)]{key-1010}
   Blain, A.W., Barnard, V.E. \& Chapman, S.C. \ 2003, MNRAS, 338, 733
\bibitem[Bolzonella et al. (2000)]{key-11}
   Bolzonella, M, Miralles, J-M, \& Pell\'o, R. \ 2000, A\&A 363, 476
\bibitem[Bouchet (1985)]{key-12}
   Bouchet, P., Lequeux, J., Maurice, E., Prevot, L., \& Prevot-Burnichon, M.L. \ 1985, A\&A, 149, 330
\bibitem[Bruzual, Charlot (1993)]{key-13}
   Bruzual, G.A., \& Charlot, S. \ 1993, ApJ, 405, 538
\bibitem[Calzetti (2000)]{key-14}
   Calzetti, D., Armus, L., Bohlin, R.C., Kinney, A.L., Koornneef, J., \& Storchi-Bergmann, T. \ 2000, ApJ, 533, 682
\bibitem[Casali, Hawarden (1992)]{key-15}
   Casali, M., \& Hawarden, T.\ 1992, UKIRT Newsletter, 4, 33
\bibitem[Chapman et al. (2002a)]{key-16}
   Chapman, S.C., Scott, D., Borys, S.C., \& Fahlman, G.G. \ 2002a, MNRAS, 330, 92
\bibitem[Chapman et al. (2002b)]{key-17}
   Chapman, S.C., Smail, I., Ivison, R.J., \& Blain, A.W. \ 2002b, MNRAS, 335, 17
\bibitem[Chapman et al. (2002c)]{key-18}
   Chapman, S.C., Shapley, A., Steidel, C.C., \& Windhorst, R. \ 2002c, ApJ, 572, L1
\bibitem[Chapman et al. (2003)]{key-19}
   Chapman, S.C., Blain, A.W., Ivison, R.J., \& Smail, I. \ 2003, Nature, 422, 695
\bibitem[Cimatti et al. (1998)]{key-20}
   Cimatti, A., Andreani, P., R\"ottgering, H., \& Tilanus, R.\ 1998, Nature, 392, 895
\bibitem[Cimatti et al. (2002)]{key-21}
   Cimatti, A., et al. \ 2002, A\&A, 381, L68
\bibitem[Daddi et al. (2000)]{key-22222222}
   Daddi, E., Cimatti, A., Pozzetti, L., Hoekstra, H., R\"ottgering, H.J.A., Renzini, A., Zamorani, G., \& Mannucci, F. \ 2000, A\&A, 361, 535
\bibitem[Daddi et al. (2001)]{key-22}
   Daddi, E., Broadhurst, T., Zamorani, G., Cimatti, A., R\"ottgering, H., \& Renzini, A.\ 2001, A\&A, 376, 825
\bibitem[Daddi et al. (2002)]{key-222222}
   Daddi, E., et al. \ 2002, A\&A, 384, L1
\bibitem[Daddi et al. (2003)]{key-2222}
   Daddi, E., et al. \ 2003, ApJ, 588, 50
\bibitem[Donahue, M. et al. (1995)]{key-23}
   Donahue, M., \& Stocke, J.T. \ 1995, ApJ, 449, 554
\bibitem[Eales et al. \ (1999)]{key-24}
   Eales, S. et al. \ 1999, ApJ, 515, 518
\bibitem[Ellingson et al. \ (1998)]{key-25}
   Ellingson, E., Yee, H.K.C., Abraham, R.G., Morris, S.L., \& Carlberg, R.G. \ 1998, ApJS, 116, 247
\bibitem[Fitzpatrick (1986)]{key-26}
   Fitzpatrick, E. L. \ 1986, AJ, 92, 1068
\bibitem[Fox et al. (2002)]{key-27}
   Fox, M.J., et al. \ 2002, MNRAS, 331, 839
\bibitem[Franx et al. (1997)]{key-28}
   Franx, M., Illingworth, G.D., Kelson, D.D., van Dokkum, P.G., \& Tran, K.-V. \ 1997, ApJ, 486, L75
\bibitem[Franx et al. (2003)]{key-2828}
   Franx, M., et al. \ 2003, ApJ, 587, L79
\bibitem[Frayer et al. (2000)]{key-29}
   Frayer, D.T., Smail, I., Ivison, R.J., \& Scoville, N.Z. \ 2000, AJ, 120, 1668
\bibitem[Frayer et al. (2003)]{key-29}
   Frayer, D.T., Armus, L., Scoville, N.Z., Blain, A.W., Reddy, N.A., Ivison, R.J., \& Smail, I. \ 2003, AJ, in press(astro-ph/0304043)
\bibitem[Fruchter, Hook (2002)]{key-30}
   Fruchter, A. S., \& Hook, R. N. \ 2002, PASP, 114, 144
\bibitem[Fukugita et al. (1996)]{key-301}
   Fukugita, M., Ichikawa, T., Gunn, J.E., Doi, M., Shimasaku, K. \& Schneider, D.P. \ 1996, AJ, 111, 1748
\bibitem[Gioia et al. (1990)]{key-31}
   Gioia, I.M., Maccacaro, T., Schild, R.E., Wolter, A., Stocke, J.T., Morris, S.L., \& Henry, J.P. \ 1990, ApJS, 72, 567
\bibitem[Gioia et al. (1998)]{key-32}
   Gioia, I.M., Shaya, E.J., Le F\'erve, O., Falco, E.E., Luppino, G.A., Hammer, F. \ 1998, ApJ 497 573
\bibitem[Holland et al. (1999)]{key-33}
   Holland, W.S., et al. \ 1999, MNRAS, 303, 659
\bibitem[Holtzman et al. (1995)]{key-34}
   Holtzman, J.A., Burrows, C.J., Casertano, S., Hester, J.J., Trauger, J.T., Watson, A.M., \& Worthey, G. \ 1995, PASP, 107, 1065
\bibitem[Hughes et al. (1998)]{key-35}
   Hughes, D.H. et al. \ 1998, Nature, 394, 241
\bibitem[Ivision et al. (2000a)]{key-36}
   Ivision, R.J., Smail. I., Barger, A.J., Kneib, J.-P., Blain, A.W., Owen, F.N., Kerr, T.H., \& Cowie, L.L. \ 2000, MNRAS, 315, 200
\bibitem[Ivision et al. (2001)]{key-38}
   Ivision, R.J., Smail, I., Frayer, D.T., Kneib, J.-P., \& Blain, A.W. \ 2001, ApJ, 561, L45
\bibitem[Ivision et al. (2002)]{key-39}
   Ivision, R.J., et al. \ 2002, MNRAS, 337, 1
\bibitem[Kashikawa et al. (2002)]{key-40}
   Kashikawa, N., et al. \ 2002, PASJ, 54, 819
\bibitem[Kneib et al. (1996)]{key-4040}
   Kneib, J.-P., Ellis, R.S., Smail, I., Couch, W.J., \& Sharples, R.M. \ 1996 ApJ 471, 643
\bibitem[Kormann et al. (1994)]{key-41}
   Kormann, R., Schneider, P., \& Bartelmann, M.\ 1994, A\&A, 284, 285
\bibitem[Landolt (1992)]{key-42}
   Landolt, A.U. \ 1992, AJ, 104, 340
\bibitem[Ledlow et al. (2002)]{key-43}
   Ledlow, M.J., Smail, I., Owen, F.N., Keel, W.C., Ivison, R.J., \& Morrison, G.E. \ 2002, ApJ, 577, L79
\bibitem[Lilly et al. (1999)]{key-44}
   Lilly, S.J., Eales, S.A., Gear, W.K.P., Hammer, F., Le F\'evre, O., Crampton, D., Bond, J.R., \& Dunne, L. \ 1999, ApJ, 518, L641
\bibitem[Lutz et al. (2001)]{key-4444}
   Lutz, D. et al. \ 2001 A\&A, 378, L70
\bibitem[Miyazaki et al. (2002)]{key-45}
   Miyazaki, S., et al. 2002, PASJ, 54, 833
\bibitem[Moustakas et al. (1997)]{key-46}
   Moustakas, L.A., Davis, M., Graham, J.R., Silk, J., Peterson, B.A., \& Yoshii, Y. \ 1997, ApJ, 475, 445
\bibitem[Motohara et al. (2002)]{key-47}
   Motohara, K., et al. \ 2002, PASJ, 54, 315
\bibitem[Pettini et al. (2000)]{key-4848}
   Pettini, M., Steidel, C.C., Adelberger, K.L., Dickinson, M., \& Giavalisco, M. \ 2000, ApJ, 528, 96
\bibitem[Prevot et al. (1984)]{key-48}
   Prevot, M.L., Lequeux, J., Prevot, L., Maurice, E., \& Rocca-Volmerange, B. \ 1984, A\&A, 132, 389
\bibitem[Pozzetti, Mannucci (2000)]{key-49}
   Pozzetti, L., \& Mannucci, F.\ 2000, MNRAS, 317, L17
\bibitem[Roche et al.(2002)]{key-50}
   Roche, N.D., Almaini, O., Dunlop, J., Ivison, R.J., \& Willot, C.J.\ 2002, MNRAS, 337, 1282
\bibitem[Schneider et al. (1992)]{key-51}
   Schneider, P., Ehlers, J., \& Falco, E.E. \ 1992, Gravitational Lenses (Berlin:Springer) p266
\bibitem[Scott et al. (2002)]{key-5252}
   Scott, S.E., et al. \ 2002, MNRAS, 331, 817
\bibitem[Shapley et al. (2001)]{key-52}
   Shapley, A., Steidel, C.C., Adelberger, K.L., Dickinson, M., Giavalisco, M., \& Pettini, M. \ 2001, ApJ, 562, 95 
\bibitem[Smail et al. (1997)]{key-5757}
   Smail, I., Ivison, R.J. \& Blain, A.W. \ 1997, ApJ, 490, L5
\bibitem[Smail et al. (1998)]{key-57}
   Smail, I., Ivison, R.J., Blain, A.W., \& Kneib, J.-P. \ 1998, ApJ, 507, L21
\bibitem[Smail et al. (1999)]{key-56}
   Smail, I., Ivison, R.J., Kneib, J.-P., Cowie, L.L., Blain, A.W., Barger, A.J., Owen, F.N., \& Morrison, G. \ 1999, MNRAS, 308, 1061
\bibitem[Smail et al. (2002a)]{key-53}
   Smail, I., Ivison, R.J., Blain, A.W. \& Kneib, J.-P. \ 2002a, MNRAS, 331, 495
\bibitem[Smail et al. (2002b)]{key-54}
   Smail, I., Owen, F.N., Morrison, G.E., Keel, W.C., Ivison, R.J. \& Ledlow, M.J. \ 2002b, ApJ, 581, 844
\bibitem[Smith et al. (2001)]{key-58}
   Smith, G.P., Treu, T., Ellis, R.S., \& Smail, I. \ 2001, ApJ, 562, 635
\bibitem[Smith et al. (2002)]{key-59}
   Smith, G.P., et al. \ 2002, MNRAS, 330, 1
\bibitem[Smith et al. (2002)]{key-60}
   Smith, J.A., et al. \ 2002, AJ, 123, 2121
\bibitem[Stanford et al. (1998)]{key-61}
   Stanford S.A., Eisenhardt, P.R., \& Dickinson, M. \ 1998, ApJ, 492, 461 
\bibitem[Stanford et al. (2002)]{key-62}
   Stanford S.A., Eisenhardt, P.R., Dickinson, M., Holden, B.P., \& De Propris, R. \ 2002, ApJS, 142, 153
\bibitem[Steidel et al. (1999)]{key-63}
   Steidel, C.C., Adelberger, K.L., Giavalisco, M., Dickinson, M., \& Pettini, M. \ 1999, ApJ, 519, 1
\bibitem[Thompson et al. (1999)]{key-64}
   Thompson, D., et al. \ 1999, ApJ, 523, 100
\bibitem[van Dokkum et al. (2003)]{key-6565}
   van Dokkum, P.G., et al. \ 2003, ApJ, 587, L83
\bibitem[Webb et al. (2002)]{key-65}
   Webb, T.M., et al. \ 2002, ApJ, 582, 6
\bibitem[Wehner et al. (2002)]{key-66}
   Wehner, E.H., Barger, A.J., \& Kneib, J.-P. \ 2002, ApJ, 577, L83
\bibitem[Waskett et al. (2003)]{key-67}
   Waskett, T.J., et al. \ 2003, MNRAS, 341, 1217
\bibitem[Williams et al. (1999)]{key-68}
   Williams, L.L.R., Navarro, J.F., \& Bartelmann, M. \ 1999, ApJ, 527, 535
\bibitem[Wu et al. (1998)]{key-69}
   Wu, X.P., Chiueh, T., Fang, L.-Z., \& Xue, Y.-J. \ 1998, MNRAS, 301, 861
\bibitem[Yagi et al. (2002)]{key-70}
   Yagi, M., Kashikawa, N., Sekiguchi, M., Doi, M., Yasuda, N., Shimasaku, K., \& Okamura, S. \ 2002, AJ, 123, 66
\bibitem[Yun \& Carilli (2002)]{key-71}
   Yun, M.S., \& Carilli, C.L. \ 2002, ApJ, 568, 88
\end{thebibliography}
\end{document}